\documentclass{aa}
\usepackage{graphicx}
\usepackage{natbib}
\bibpunct{(}{)}{;}{a}{}{,}

\begin{document}
\title{The low-luminosity tail of the GRB distribution\,:\\ the case of GRB 980425.}
\author{F. Daigne\inst{1,2} \and R. Mochkovitch\inst{1}}
\institute{Institut d'Astrophysique de Paris, 98 bis boulevard Arago, 75014 Paris France
\and Universit\'{e} Pierre et Marie Curie-Paris VI, 4 place Jussieu, 75005 Paris, France}
\offprints{F. Daigne, \email{daigne@iap.fr}}
\date{Received 21 July 2006 / Accepted 23 November 2006}
\abstract{The association of GRB 980425 with the nearby supernova SN 1998bw at z=0.0085 implies the existence of a population of gamma-ray bursts with an isotropic-equivalent luminosity which is about $10^{4}$ times smaller than in the standard cosmological case. Apart from its weak luminosity, GRB 980425 appears as a ``normal'' burst regarding all its other properties (variability, duration, spectrum), with however a rather low peak energy $E_\mathrm{p} \simeq 30-100\ \mathrm{keV}$.}
{We investigate two scenarios to explain a weak gamma-ray burst such as GRB 980425 : a normal (intrinsically bright) gamma-ray burst seen off-axis or an intrinsically weak gamma-ray burst seen on-axis.}
{For each of these two scenarios, we first derive the conditions to produce a GRB 980425-like event and we then discuss the consequences for the event rate. In the second scenario, this study is done in the framework of the internal shock model.}
{If we exclude the possibility that GRB 980425 is an occurence of an extremely rare event observed by chance during the first eight years of the ``afterglow era'', 
the first scenario implies that (i) the local rate of ``standard'' bright gamma-ray bursts is much higher than what is usually expected; 
(ii) the typical opening angle in gamma-ray bursts ejecta is much narrower than what is derived from observations of a break in the afterglow lightcurve.
In addition to this statistical problem, we show that the afterglow of GRB 980425 in this scenario should have been very bright and easily detected.
For these reasons the second scenario appears more realistic. We show that the parameter space of the internal shock model indeed allows GRB 980425-like events, in cases where the outflow is only mildly-relativistic and mildly-energetic. The rate of such weak events in the Universe has to be much higher than the rate of ``standard'' bright gamma-ray bursts to allow the discovery of GRB 980425 during a short period of a few years. However it is still compatible with the observations as the intrinsic weakness of these GRB 980425-like bursts does not allow detection at cosmological redshift with present gamma-ray instruments.
We finally briefly discuss the consequences of such a high local rate of GRB 980425-like events for the future prospects of detecting non-electromagnetic radiation, especially gravitational waves.}
{}

\maketitle

\section{Introduction}

Our knowledge of the gamma-ray burst (hereafter GRB) population has dramatically increased in the recent years. In addition to ``standard'' cosmological bright GRBs peaking at 100 keV -- 1 MeV have been discoved X-ray Flashes (hereafter XRFs) and X-ray Rich GRBs (hereafter XRRs) that peak at much lower energy \citep{heise:01,kippen:01,barraud:03}, weak GRBs \citep[such as GRB 031203,][]{sazonov:04,soderberg:04} and even a few ``local'' events: GRB 980425 \citep{galama:98} and GRB 060218 \citep{mirabal:06}. It is usually believed that a long GRB is produced by an ultra-relativistic outflow ejected from a black hole that has just formed in the collapse of a very massive star (collapsars, \citet{woosley:93}). The prompt GRB emission originates from the relativistic ejecta itself, and is probably due to the formation of internal shocks \citep{rees:94}, whereas the afterglow is emitted by the strong shock that propagates within the ambient medium during the deceleration phase of the outflow \citep{meszaros:97}.\\
It is important to understand the physical origin of the observed diversity in this theoretical framework. There are two groups of explanations : either the diversity is only apparent: all GRBs are intrinsically very similar, but the observation conditions (viewing angle for instance) break this similarity; or the diversity is intrinsic: all collapsars do not produce the same relativistic outflow and therefore the same GRB.\\
In this paper, we focus on the closest GRB, i.e. GRB 980425. This burst has been detected by \textit{Beppo-SAX} and \textit{BATSE}. It appears as a ``standard'' single-pulse burst regarding its gamma-ray properties : duration $T_{\gamma}\sim 31\ \mathrm{s}$, peak flux (40-700 KeV) $P\sim 2.4\times 10^{-7}\ \mathrm{erg/cm^{2}/s}$, Band spectrum with low- and high-energy slopes $\alpha\sim -0.8$ and $\beta \sim -2.3$ and peak energy $E_\mathrm{p}\sim 68\pm 40~\mathrm{keV}$ \citep{frontera:00}. With maybe the exception of a somewhat lower peak energy, all these properties are very close to those of standard GRBs. Note also that \citet{norris:00} derived an especially large time lag for this burst between \textit{BATSE} channels 1 and 3 : $\Delta_{13}\sim 4.5\ \mathrm{s}$.\\
A peculiar type Ic supernova, SN 1998bw,  was discovered simultaneously in the \textit{Beppo-SAX} error-box \citep{galama:98}. The probability to have two such rare events occuring at the same time in the same direction is very low, so that GRB 980425 and SN 1998bw are probably physically associated \citep[see also][]{kouveliotou:04}. This is reinforced by other associations that have been found since 1998, the better case being probably the association of GRB 030329 with SN2003dh \citep{stanek:03} at $z=0.168$. If the association GRB 980425 / SN 1998bw is real, then this burst is far from ``standard''. The host galaxy of SN 1998bw is indeed located at $z=0.0085$ and GRB 980425 is therefore the closest GRB ever detected. As the peak flux of GRB 980425 is comparable to typical peak fluxes of other \textit{Beppo-SAX} GRBs, it means that it is intrinsically much weaker (by more than four orders of magnitude).\\
In section~\ref{sec:peakflux}, we recall the various physical factors entering in the observed peak flux and peak energy of a cosmic GRB. We then study in section~\ref{sec:openingangle} a scenario where GRB 980425 is a standard bright GRB seen off-axis, and therefore apparently weak. In section~\ref{sec:is} we detail an alternative explanation, where GRB 980425 is intrinsically weak. We show how the internal shock model can allow for such weak bursts. Both in section~\ref{sec:openingangle} and \ref{sec:is} we discuss the consequences of each scenario in terms of GRB rate. Our results are summarized in section~\ref{sec:conclusions}.


\section{Peak flux and peak energy of a cosmic GRB}
\label{sec:peakflux}

\begin{figure}
\centering
\resizebox{\hsize}{!}{\includegraphics*[4cm,10cm][21cm,17.5cm]{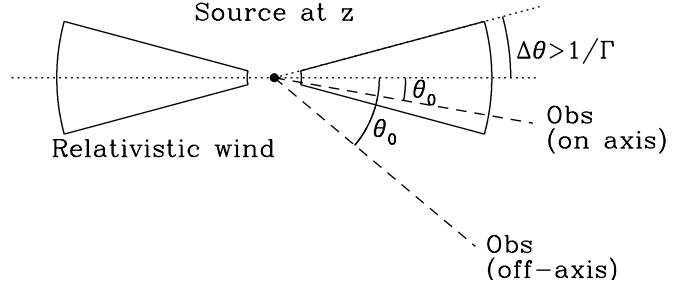}}\\
\caption{\textbf{Geometry.} A relativistic outflow of Lorentz factor $\Gamma$ is emitted by the source (located at redshift $z$) in a cone of opening angle $\Delta\theta$. The line-of-sight of the observer makes an angle $\theta_{0}$ with the axis of the cone. The observation is made on-axis (resp. off-axis) if $\theta_{0} \le \Delta\theta$ (resp. $\theta_{0} > \Delta\theta$).}
\label{fig:picgrb}
\end{figure}

We assume that a GRB is produced by a relativistic outflow of Lorentz factor $\Gamma$ and opening angle $\Delta\theta\gg 1/\Gamma$ generated by a source at redshift $z$ (see Figure~\ref{fig:picgrb}). We define $\theta_{0}$ as the angle between the line-of-sight and the axis of the ejecta. The observed bolometric peak flux and peak energy are given by
\begin{equation}
P^\mathrm{obs}  =  \mathcal{K}_\mathrm{P}\left(\Gamma; \Delta\theta; \theta_{0}\right)\times \frac{L_{\mathrm{rad},4\pi}}{4\pi D_\mathrm{L}^{2}} 
\label{eq:pobs}
\end{equation}
and
\begin{equation}
E_{p}^\mathrm{obs} = \mathcal{K}_\mathrm{E}\left(\Gamma; \Delta\theta; \theta_{0}\right)\times \frac{E_{p}}{1+z}\ ,
\label{eq:epobs}
\end{equation}
where $L_{\mathrm{rad},4\pi}$ and $E_\mathrm{p}$ are 
the isotropic equivalent luminosity and the peak energy measured in the GRB source frame. The correction for the viewing angle are approximatively given in the on-axis and the off-axis cases by 
\begin{equation}
\mathcal{K}_\mathrm{P}\left(\Gamma; \Delta\theta; \theta_{0}\right) = 
\left\lbrace\begin{array}{cl}
1 & \mathrm{if}\ \theta_{0} < \Delta\theta\\
\frac{1}{2\left(1+\Gamma^{2}\left(\theta_{0}-\Delta\theta\right)^{2}\right)^{3}} & \mathrm{if}\ \theta_{0}>\Delta\theta
\end{array}\right.
\label{eq:kp}
\end{equation}
and
\begin{equation}
\mathcal{K}_\mathrm{E}\left(\Gamma; \Delta\theta; \theta_{0}\right) = 
\left\lbrace\begin{array}{cl}
1 & \mathrm{if}\ \theta_{0} < \Delta\theta\\
\frac{1}{1+\Gamma^{2}\left(\theta_{0}-\Delta\theta\right)^{2}} & \mathrm{if}\ \theta_{0}>\Delta\theta
\end{array}\right.\ .
\label{eq:kep}
\end{equation}
Therefore, the peculiar properties of GRB 980425 may have two origins : (i) either GRB 980425 is a ``standard'' GRB, i.e. intrinsically bright with
$L_\mathrm{rad,4\pi}\ga \mathrm{a\ few}\ 10^{50}\ \mathrm{erg.s^{-1}}$
, but appears as a weak GRB because it is seen with a large viewing angle; (ii) or GRB 980425 is an intrinsically weak GRB ($L_\mathrm{rad,4\pi}\simeq 3\times 10^{46}\ \mathrm{erg.s^{-1}}$) seen on-axis. In this case GRB 980425 has been detected only because of its very low redshift.
In the following, we successively investigate these two scenarios. The first one has already been adressed by several authors \citep[e.g.][]{nakamura:99,salmonson:01,yamazaki:03,guetta:04} so we focus mainly on the second scenario. For this case, the intrinsic properties of GRBs are studied in the framework of the internal shock model, where
the prompt emission comes 
from a relativistic wind which converts part of its kinetic
energy into radiation via the formation of shock waves within
the wind itself. Such internal shocks can occur if the wind
is generated with a highly variable Lorentz
factor \citep{rees:94}. 


\begin{figure}
\centering
\resizebox{\hsize}{!}{\includegraphics{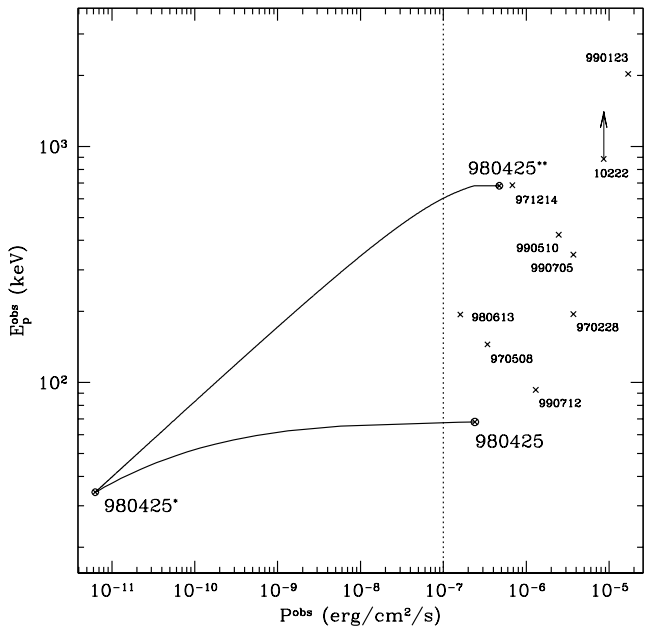}}
\caption{\textbf{Bolometric peak flux vs peak energy diagram\,:} All \textit{Beppo-SAX} GRBs with a known redshift are indicated in this diagram (crosses). The peculiar burst GRB 980425 is indicated by a circled cross. 
A vertical dotted line stands for a threshold $P_\mathrm{min}=10^{-7}\ \mathrm{erg~cm^{-2}~s^{-1}}$ representative of the \textit{BATSE} and \textit{Beppo-SAX WFC+GRBM} instruments \citep{band:03}.
\textit{Effect of the redshift~:} we have plotted the evolution of GRB 980425 when its redshift increases from $z=0.008$ to $z=1$ (cosmological distance). The final result is indicated as GRB 980425$^{*}$. \textit{Effect of the viewing angle~:} we have then plotted the evolution of GRB 980425$^{*}$ when the viewing angle decreases. We assume that GRB 980425$^{*}$ is seen off-axis with $\theta_{0}=\Delta\theta+4/\Gamma$ (see text). The final result for $\theta_{0}=0$ (on-axis) is indicated as GRB 980425$^{**}$.}
\label{fig:anglevsz}
\end{figure}


\begin{figure}
\centering
\resizebox{\hsize}{!}{\includegraphics{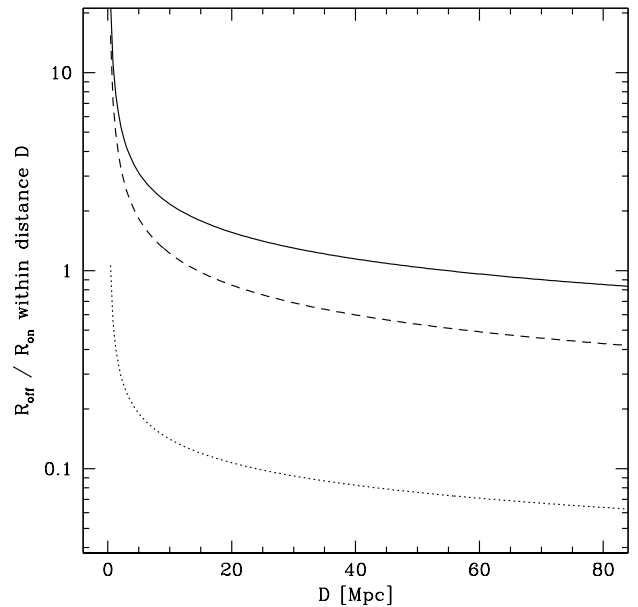}}
\caption{\textbf{The ratio of detected off-axis to on-axis GRBs in the local Universe~:} this ratio within a distance $D$ is plotted as a function of $D$ for a power-law luminosity function of slope $-1.7$ between $L_\mathrm{min}=10^{50}\ \mathrm{erg~s^{-1}}$ and $L_\mathrm{max}=10^{54}\ \mathrm{erg~s^{-1}}$ and three different distributions of the opening angle\,: (1) a uniformly distributed opening angle between $1$ and $10\degr$ (solid line); (2) a power-law distribution $p(\Delta\theta)\propto \Delta\theta^{-2}$ between $1/\Gamma$ and $\pi/2$ (dashed line); (3) an opening angle correlated with the isotropic luminosity (dotted line). In this last case, the true luminosity $L\left(1-\cos{\Delta\theta}\right)$ is assumed to be constant and equal to $L_\mathrm{min}$.}
\label{fig:stat}
\end{figure}


\section{GRB 980425 as an intrinsically bright GRB seen off-axis}
\label{sec:openingangle}
\subsection{Conjugate effect of a low redshift and a large viewing angle}
In Figure~\ref{fig:anglevsz}, GRB 980425 has been plotted in a bolometric peak flux vs peak energy diagram, together with the GRBs detected by \textit{Beppo-SAX} which have a known redshift. A sensitivity of $10^{-7}\ \mathrm{erg~cm^{-2}~s^{-1}}$ representative of the \textit{BATSE} and \textit{Beppo-SAX GRBM+WFC} instruments is also indicated as a dotted line \citep{band:03}.
We use equations~(\ref{eq:pobs})--(\ref{eq:kep}) to estimate under which conditions on the redshift and the viewing angle GRB 980425 can be an intrinsically bright GRB despite its low apparent intensity. To clarify the respective roles of redshift and viewing angle, we perform the following two transformations\,:\\ 
\noindent-- \textit{Effect of the redshift :} GRB 980425 has first been moved from $z=0.008$ to $z=1$ (cosmological distance). Both the peak energy and the peak flux decrease due to the increase of redshift and luminosity distance.
The final result is named GRB 980425$^{*}$ : this burst is clearly much to weak to be observed as a GRB or even a XRF. Its bolometric peak flux is indeed at least three orders of magnitude below the sensitivity of past or current detectors.\\
\noindent-- \textit{Effect of the viewing angle :} the viewing angle of GRB 980425$^{*}$ is then decreased down to $\theta_{0}=0$ (on-axis observation), assuming that GRB 980425 is initially seen off-axis with a large viewing angle $\theta_{0}=\Delta\theta+{k}/{\Gamma}$ (here $k=4$). This evolution is computed from the approximate formulae~(\ref{eq:kp}) and (\ref{eq:kep}). The final result is named GRB 980425$^{**}$. Now this burst is clearly back in the ``standard'' GRB region, very close to GRB 971214.\\
Therefore, as it was already pointed out by \citet{yamazaki:03}, the peculiar properties of GRB 980425 are compatible with those of an intrinsically bright GRB observed off-axis. In the case illustrated in Figure~\ref{fig:anglevsz}, the decrease of the flux of GRB 980425 by a factor $\sim 10^{4}$ due to a viewing angle $\theta_{0}=\Delta\theta+4/\Gamma$ is compensated by a smaller luminosity distance $\left(D_\mathrm{L}(z=0.008) / D_\mathrm{L}(z=1)\right)^{2} \sim 2\times 10^{-4}$.\\

If it is clear from the previous analysis that a slightly off-axis viewing angle ($\theta_{0}-\Delta\theta\sim 4/\Gamma\sim 2.3\degr$ if $\Gamma\simeq 100$) can account for the properties of GRB 980425, one should however keep in mind that the peak flux decreases very rapidly with viewing angle (see equation~(\ref{eq:kp})). 
For $\theta_{0}=\Delta\theta+k/\Gamma$ with $k\ge 6$, the peak flux is divided by a factor larger than $\sim 10^{5}$ and the decrease of luminosity distance is not able anymore to compensate\,: such off-axis GRBs cannot be detected. As $\Gamma \ga 100$ in GRB outflows \citep{baring:95,lithwick:01}, the angle $6/\Gamma \la 3.4\degr$ is small and the probability to observe GRB 980425-like events should be low.

\subsection{A statistical problem ?}

For a detector having a threshold $P_\mathrm{min}$ (hereafter we assume $P_\mathrm{min}=10^{-7}\ \mathrm{erg~cm^{-2}~s^{-1}}$, which is representative of instruments such as \textit{Beppo-SAX GRBM+WFC} or \textit{BATSE}) the observed rate of on-axis GRBs within $D=40\ \mathrm{Mpc}$ is given by
\begin{equation}
\mathcal{R}_\mathrm{on} = \frac{4\pi}{3} D^{3} R_{0} \times \int_{0}^{\pi/2} \mathrm{d}\Delta\theta~p\left(\Delta\theta\right)\left(1-\cos{\Delta\theta}\right)\ ,
\end{equation}
where $R_{0}$ is the local GRB volumic rate  and $p(\Delta\theta)$ is the probability distribution of the opening angle $\Delta\theta$. Here we assume that the minimum GRB luminosity $L_\mathrm{min}$ is larger than $4\pi D^{2} P_\mathrm{min}=2\times 10^{46}\ \mathrm{erg~s^{-1}}$ so that all nearby on-axis GRBs are detected. On the other hand, the rate of off-axis GRBs in the same volume is affected by the rapid decrease of the GRB flux with the opening angle so that 
\begin{eqnarray}
\mathcal{R}_\mathrm{off} & = & R_{0} \int_{0}^{D} \mathrm{d}D~4\pi D^{2} 
\int_{0}^{\pi/2} \mathrm{d}\Delta\theta~p\left(\Delta\theta\right) 
\int_{L_\mathrm{min}}^{L_\mathrm{max}} \mathrm{d}L~p(L) \nonumber\\
& &
\times \left(\cos{\Delta\theta}-\cos{\left(\Delta\theta+\frac{k_\mathrm{max}\left(D,L\right)}{\Gamma}\right)}\right)\ ,
\label{eq:ron}
\end{eqnarray}
where $p(L)$ is the GRB luminosity function and $k_\mathrm{max}$ is given by
\begin{equation}
k_\mathrm{max}\left(D,L\right) \simeq \sqrt{\left(\frac{1}{2}~\frac{L}{4\pi D^{2} P_\mathrm{min}}\right)^{1/3}-1}\ .
\label{eq:roff}
\end{equation}
In both equations~(\ref{eq:ron}) and (\ref{eq:roff}), the probability distribution of the Lorentz factor $\Gamma$ is neglected and $\Gamma$ is supposed to be constant. In the most simple case where both the luminosity $L$ and the opening angle $\Delta\theta$ are constant, we find that the ratio of detected off-axis over on-axis GRBs can be expressed as a ratio of two solid angles :
\begin{equation}
\frac{\mathcal{R}_\mathrm{off}}{\mathcal{R}_\mathrm{on}} \simeq \frac{\cos{\Delta\theta}-\cos{\left(\Delta\theta+\frac{k_\mathrm{max}\left(D,L\right)}{\Gamma}\right)}}{1-\cos{\Delta\theta}}\ .
\label{eq:ratio}
\end{equation}
For $D=40\ \mathrm{Mpc}$ 
and $L=10^{51}\ \mathrm{erg~s^{-1}}$, we get $k_\mathrm{max}=5.4$,  
so that $k_\mathrm{max}/\Gamma \la 3.1\degr$ 
for $\Gamma \ga 100$. This leads to a rate of detected off-axis over on-axis GRBs
$\mathcal{R}_\mathrm{off}/\mathcal{R}_\mathrm{on}\sim 16$ 
for $\Delta\theta=1\degr$,
$\mathcal{R}_\mathrm{off}/\mathcal{R}_\mathrm{on}\sim 1$
for $\Delta\theta=7.5\degr$,
 and $\mathcal{R}_\mathrm{off}/\mathcal{R}_\mathrm{on}\sim 0.7$ 
for $\Delta\theta=10\degr$.
This implies that the rates of detected off-axis and on-axis GRBs should be comparable for typical $\Delta\theta$ of a few degrees and that it is only for $\Delta\theta \la 1\degr$ that the number of detected off-axis GRBs is much larger than the number of on-axis GRBs.
The local apparent rate of ``standard'' on-axis bright GRBs can be estimated to be of the order of $\sim 1 / \left(5000-30000\ \mathrm{yr}\right)$ within 40 Mpc \citep{porciani:01,schmidt:01,perna:03,guetta:04}. This simple analysis would then imply that with GRB 980425 we have observed by chance a very rare event.
Even if it has some rather different properties, GRB 060218/SN2006aj at $z=0.0331$ (144 Mpc) \citep{cusumano:06,mirabal:06,masetti:06,soderberg:06}) is another case of a nearby burst. At this distance $\mathcal{R}_\mathrm{off}/\mathcal{R}_\mathrm{on}$ should be even smaller. Statistical studies \citep{bosnjak:06} also indicate that there are probably other GRB980425-like events in the \textit{BATSE} catalog. For all these reasons, we conclude that the off-axis interpretation probably suffers a statistical problem. 
In the more general case, the two solid angles in equation~(\ref{eq:ratio}) have to be averaged over the luminosity function, the distance and the opening angle distribution, according to equations~(\ref{eq:ron}) and (\ref{eq:roff}). We did that for a power-law luminosity function with slope $-1.7$ \citep[this slope giving a good fit to the $\log{N}-\log{P}$ diagram, see e.g][]{daigne:06}
between $L_\mathrm{min}=10^{50}~\mathrm{erg~s^{-1}}$ and  $L_\mathrm{max}=10^{54}~\mathrm{erg~s^{-1}}$. We tested three possible distributions for the opening angle\,: (1) a uniformly distributed opening angle; (2) a power-law distributed opening angle; (3) an opening angle correlated with the isotropic equivalent luminosity so that the true luminosity is constant, as suggested by observations of achromatic breaks in afterglow lightcurves \citep{frail:01}. We find (see Figure~\ref{fig:stat}) that except below $\sim 5~\mathrm{Mpc}$ where the total (on+off axis) event rate is very low due to a small volume, the expected rate of off-axis GRBs is either comparable (uniformly distributed opening angle) or smaller than the on-axis GRB rate. This is in full agreement with our simple estimate (equation~(\ref{eq:ratio})). This equation also indicates that the only way to have a much higher rate of off-axis GRBs is to assume a very small opening angle ($\Delta\theta \ll 1/\Gamma$), in contradiction with observations \citep{frail:01}.


\section{GRB 980425 as an intrinsically weak GRB}
\label{sec:is}

We now consider in this section an alternative scenario where GRB 980425 is an intrinsically weak GRB seen on-axis. The intrinsic GRB properties are considered in the framework of the internal shock model \citep{rees:94}.
\subsection{Internal shocks}
The dynamics of the internal shocks as well as the temporal and spectral properties of the emission have been studied in details in \citet{daigne:98} using a simplified model where the outflow is made of a large number of discrete relativistic shells that interact by direct collision only. This approach has been validated by 1D relativistic hydrodynamical simulations \citep{daigne:00}. We adopt here an even simpler version of the model, where we consider only a typical internal shock due to the collision of two shells of equal mass $M$ and Lorentz factors $\Gamma_{1}$ and $\Gamma_{2}>\Gamma_{1}$. This toy model is described in \citet{barraud:05}. It is found that the collision between the two shells occurs at radius
\begin{equation}
R_\mathrm{is} \simeq \frac{8\kappa^{2}}{(\kappa-1)(\kappa+1)^{3}} \Gamma^{2} c\tau\ ,
\end{equation}
that the isotropic equivalent radiated luminosity due to internal shocks is given by
\begin{equation}
L_\mathrm{rad,4\pi} \simeq \epsilon_\mathrm{e} \frac{(\sqrt{\kappa}-1)^{2}}{\kappa+1} \dot{E}
\label{eq:lrad}
\end{equation}
and that the peak energy takes the form
\begin{equation}
E_\mathrm{p} \simeq K \frac{\dot{E}^{x}~\phi_{xy}(\kappa)}{\tau^{2x}~\Gamma^{6x-1}}\ ,
\label{eq:ep}
\end{equation}
where
$\dot{E}$ is the isotropic equivalent kinetic energy injection rate in the relativistic outflow,
$\tau$ is the duration of the relativistic ejection (i.e. here the time interval between the two shell ejections), $\Gamma$ is the mean Lorentz factor $(\Gamma_{1}+\Gamma_{2})/2$ and the contrast $\kappa=\Gamma_{2}/\Gamma_{1}$ is a measure of the initial amplitude of the variations of the Lorentz factor in the ejecta; $\epsilon_\mathrm{e}$ is the fraction of the dissipated energy in the shock that is injected into non-thermal electrons, and $K$ is a constant.
The function $\phi_{xy}(\kappa)$ is given by
\begin{equation}
\phi_{xy}(\kappa) = \frac{\left(\sqrt{\kappa}-1\right)^{2y}\left(\kappa-1\right)^{2x}\left(\kappa+1\right)^{6x-1}}{\kappa^{2x+\frac{y-1}{2}}}
\end{equation}
and is steadily increasing with $\kappa$. The $x$ and $y$ parameters have been introduced by considering that the peak energy in the comoving frame of the shocked material scales as 
\begin{equation}
E_\mathrm{p}' \propto \rho^{x} \epsilon^{y}\ ,
\end{equation}
where $\rho$ is the comoving density and $\epsilon$ the dissipated energy per unit mass.\\

In the standard synchrotron model with constant equipartition  parameters it is assumed that the magnetic field is amplified to reach a fraction $\epsilon_\mathrm{B}$ of the total dissipated energy in the shock, and that a fraction $\epsilon_\mathrm{e}$ of this dissipated energy is injected into a fraction $\zeta$ of the electrons, which are therefore accelerated to high Lorentz factors.
This leads to $x=1/2$ and $y=5/2$, as the synchrotron energy scales as $E'_\mathrm{p}\propto B \Gamma_\mathrm{e}^{2}$, the magnetic field as $B\propto \epsilon_\mathrm{B}^{1/2} (\rho\epsilon)^{1/2}$ and the typical Lorentz factor of the electrons as $\Gamma_\mathrm{e}\propto (\epsilon_\mathrm{e}/\zeta) \epsilon$. However \citet{daigne:03} have shown that smaller values of $x$ and $y$ are required to reproduce the hardness-intensity and hardness-fluence correlations observed in many GRB pulses \citep{golenetskii:83,liang:96}. They consider $x=y=0.5$ and $x=y=0.25$. Such exponents can for example be obtained with the standard synchrotron process if the equipartition parameters vary with shock intensity.

\subsection{Producing GRB980425-like events}


\begin{figure*}
\centering
\resizebox{\hsize}{!}{\includegraphics{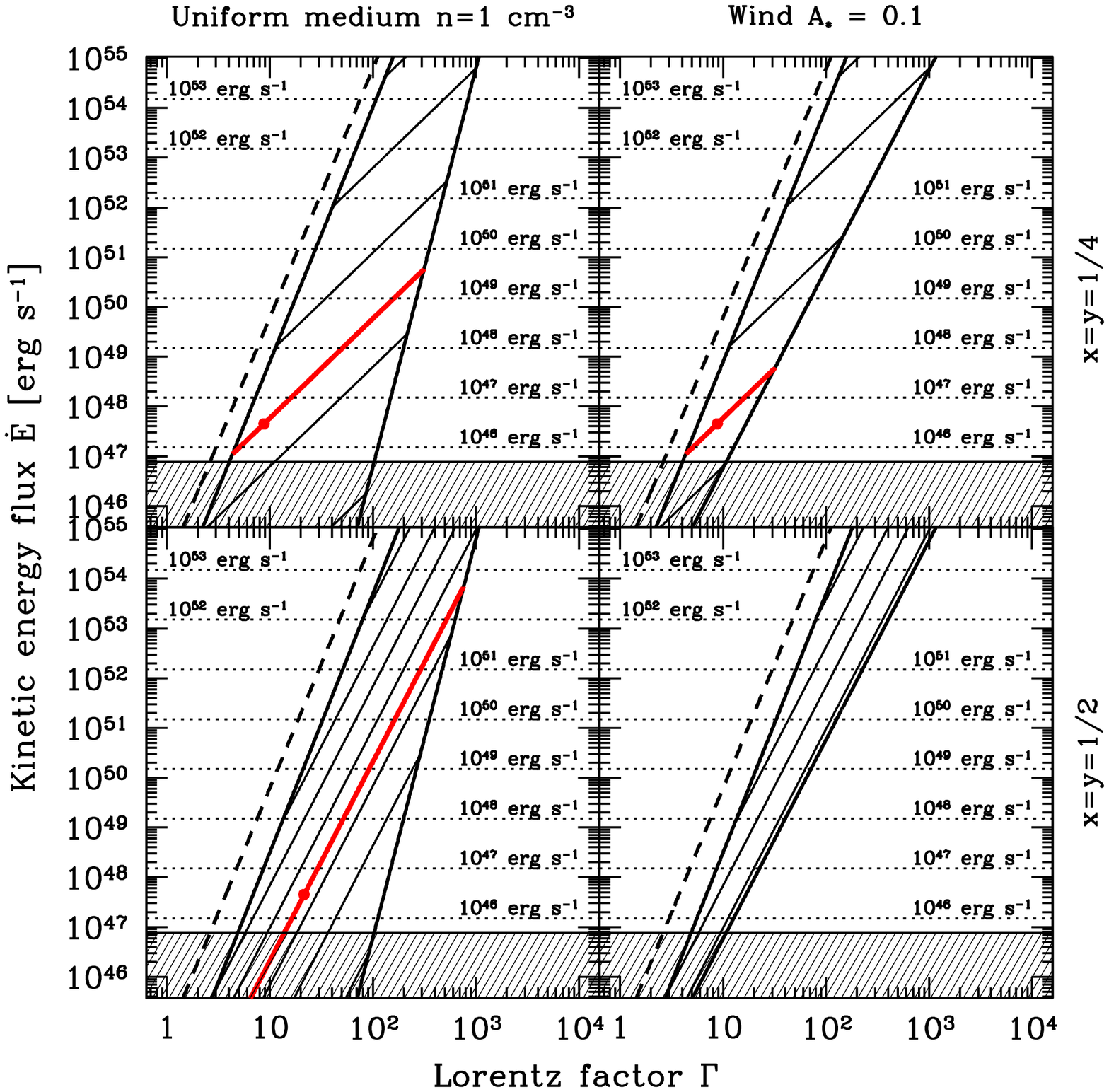}}
\caption{\textbf{The internal shock model parameter space at $z=0.0085$}. In the $\Gamma$--$\dot{E}$ plane, lines of constant peak energy (observer frame) are plotted for $\kappa=4$, $\tau=30\ \mathrm{s}$, $\epsilon_\mathrm{e}=1/3$ and a fixed value of $K$ computed as explained in the text. These lines are limited on the left side by the transparency constraint given by equation~(\ref{eq:c2}) (optical depth for pair creation~: solid line) and (\ref{eq:c1}) (opacity due to ambient electrons~: dashed line) and on the right side by the environment constraint given by equation~(\ref{eq:c3}). As $\kappa$ and $\epsilon_\mathrm{e}$ are fixed, the radiative efficiency is also fixed to 6.7\% 
(see text)~: thin dotted horizontal lines show the radiated luminosity $L_\mathrm{rad,4\pi}$ corresponding to the injected kinetic energy rate $\dot{E}$. The shaded area excludes GRBs that cannot be observed by a bolometric detector of threshold $P_\mathrm{min}=10^{-7}~\mathrm{erg~cm^{-2}~s^{-1}}$. The thick line corresponds to $E_\mathrm{p}=70~\mathrm{keV}$. Above this line are also plotted lines for $E_\mathrm{p}=150~\mathrm{keV}$, $400~\mathrm{keV}$, $1~\mathrm{MeV}$, and $3~\mathrm{MeV}$. Below are found $E_\mathrm{p}=40~\mathrm{keV}$ and $10~\mathrm{keV}$. The location of GRB980425 is shown as a big dot.
}
\label{fig:isparameters1}
\end{figure*}

The value of $L_{\mathrm{rad},4\pi}$ and $E_\mathrm{p}$ given by equations~(\ref{eq:lrad}) and (\ref{eq:ep}) are fixed by 6 physical quantities~: $\dot{E}$, $\bar{\Gamma}$, $\kappa$, $\tau$, $\epsilon_\mathrm{e}$ and $K$. The value of $\epsilon_\mathrm{e}$ and $K$ depend on the details of the physical processes in the shocked material, which are not studied here. We adopt $\epsilon_\mathrm{e}=1/3$ and we fix the value of $K$ by demanding that a ``typical'' GRB at redshift $z=1$ with $\tau=5\ \mathrm{s}$ (observed duration of $(1+z)\tau=10\ \mathrm{s}$), $\kappa=4$, $\dot{E}=1.5\times 10^{52}\ \mathrm{erg.s^{-1}}$ ($L_\mathrm{rad,4\pi}=10^{51}\ \mathrm{erg.s^{-1}}$) and $\bar{\Gamma}=300$ has an observed peak energy $E_\mathrm{p}^\mathrm{obs}=200\ \mathrm{keV}$. Note that in this case, the efficiency of internal shocks is $L_\mathrm{rad,4\pi}/\dot{E}=6.7\%$, as the product of $\epsilon_\mathrm{e}=1/3$ with an internal shock dynamical efficiency of $20\%$
for $\kappa=4$ (see equation~\ref{eq:lrad}).
The other parameters are limited by several constraints \citep{daigne:04}\,:

\paragraph{\textbf{(1) Transparency during the internal shock phase.}}
The relativistic ejecta has to be transparent during the internal shock phase. The Thomson optical depth of the outflow is given by \citep{meszaros:00,daigne:02}~:
\begin{equation}
\tau(R_\mathrm{is}) \simeq \frac{\sigma_{T}\dot{E}\tau}{4\pi \Gamma R_\mathrm{is} \left(R_\mathrm{is}+2\Gamma^{2}c\tau\right) m_\mathrm{p} c^{2}}\ .
\label{eq:ris}
\end{equation}
Then the condition  $\tau(R_\mathrm{is})<1$ leads to
\begin{equation}
\frac{\dot{E}}{\Gamma^{5}} < \frac{4\pi m_\mathrm{p} c^{4}}{\sigma_\mathrm{T}}\ f_{1}(\kappa)\ \tau\  ,
\label{eq:c1}
\end{equation}
with
\begin{equation}
f_{1}(\kappa) = \frac{8\kappa^{2}}{(\kappa-1)(\kappa+1)^{3}} \left( \frac{8\kappa^{2}}{(\kappa-1)(\kappa+1)^{3}} + 2 \right)\ .
\end{equation}

\paragraph{\textbf{(2) No pair production.}}
The relativistic ejecta has to be transparent to pairs during the internal shock phase. Pairs are produced by photon-photon annihilation in the high-energy part of the GRB spectrum. 
At radius $R_\mathrm{is}$, the approximate optical depth for pair creation is \citep{meszaros:00,lithwick:01} :
\begin{equation}
\tau_{\pm}(R_\mathrm{is}) \simeq \left[ \alpha_{\pm} \frac{\sigma_\mathrm{T} L_\mathrm{rad,4\pi}\tau}{4\pi R^{2} \Gamma m_\mathrm{e} c^{2}} \left(\frac{E_\mathrm{p}}{\Gamma m_\mathrm{e}c^{2}}\right)^{\beta-2}\right]^{2}\ ,
\end{equation}
where $\beta$ is the high-energy slope of the spectrum and $\alpha_{\pm}$ is a dimensionless number that depends on the spectral shape. In the following, we adopt $\beta=2.5$ \citep{preece:00}. This leads to $\alpha_{\pm}\simeq 0.06$. Then the condition $\tau_{\pm}(R_\mathrm{is})<1$ leads to
\begin{equation}
\frac{\dot{E}^{1+x(\beta-2)}}{\Gamma^{5+6x(\beta-2)}} < \frac{4\pi m_{e}c^{4}}{\alpha_{\pm}\sigma_\mathrm{T}} \frac{f_{2}(\kappa)}{\epsilon_\mathrm{e}} \left(\frac{K}{m_\mathrm{e}c^{2}}\right)^{2-\beta} \tau^{1+2x(\beta-2)}
\label{eq:c2}
\end{equation}
with
\begin{equation}
f_{2}(\kappa) = \frac{64 \kappa^{4}}{(\sqrt{\kappa}-1)^{2}(\kappa-1)^{2}(\kappa+1)^{5}~\Phi_{xy}^{\beta-2}(\kappa)}
\end{equation}
We also considered an additional constraint provided by the lack of an observed high-energy cutoff due to photon-photon annihilation in GRB spectra \citep[see e.g.][]{lithwick:01}. However, the constraint expressed by equation~(\ref{eq:c2}) is dominant in most cases;

\paragraph{\textbf{(3) Internal shocks operate before the deceleration phase.}}
The internal shock phase has to occur before the deceleration of the ejecta  due to the external medium.  For a density $\rho \propto A / R^{s}$ in the environment, the deceleration starts at
\begin{equation}
R_\mathrm{dec} \simeq \left(\frac{3-s}{4\pi}\ \frac{\dot{E}\tau}{A\Gamma^{2}c^{2}}\right)^{\frac{1}{3-s}}\ .
\end{equation}
Then the condition $R_\mathrm{is} < R_\mathrm{dec}$ leads to
\begin{equation}
\frac{\dot{E}}{\Gamma^{8-2s}} > \frac{4\pi}{3-s} A c^{5-s} \tau^{2-s} 
\left(\frac{8\kappa^{2}}{(\kappa-1)(\kappa+1)^{3}}\right)^{3-s}\ .
\label{eq:c3}
\end{equation}


\begin{figure}
\centering
\resizebox{\hsize}{!}{\includegraphics{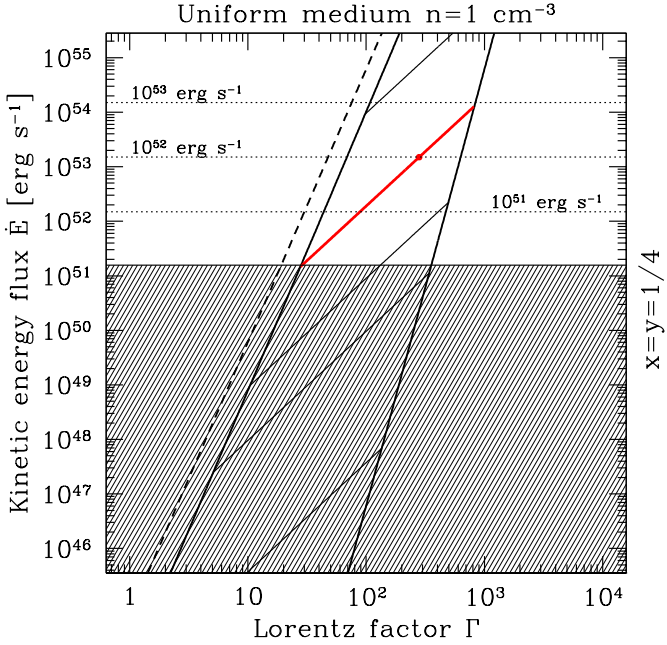}}
\caption{\textbf{The internal shock model parameter space at $z=1$}. The diagram plotted in Figure~\ref{fig:isparameters1} has been moved from $z=0.0085$ to $z=1$, $\kappa$, $\tau$, $\epsilon_\mathrm{e}$ and $K$ being kept constant. Only the case where $x=y=1/4$ and the external medium has a constant density $n=1\ \mathrm{cm^{-3}}$ is shown. Weak GRBs are now undetected. The thick line corresponds to $E_\mathrm{p}=150~\mathrm{keV}$. Above this line are found $E_\mathrm{p}=400\ \mathrm{keV}$ and $1\ \mathrm{MeV}$. Below are plotted lines for $E_\mathrm{p}=70\ \mathrm{keV}$, $40\ \mathrm{keV}$ and $10\ \mathrm{keV}$. The big dot indicates the location of a ``standard'' GRB with $E_\mathrm{p}=150~\mathrm{keV}$ (observer frame) and $L_\mathrm{rad,4\pi}=10^{52}\ \mathrm{erg~s^{-1}}$.}
\label{fig:isparameters2}
\end{figure}

\vspace*{1ex}

Figure~\ref{fig:isparameters1} shows the location of these constraints as well as lines of constant peak energy in the $\Gamma$--$\dot{E}$ plane for two sets of the $x$ and $y$ parameters : $x=y=1/4$ (top) and $x=y=1/2$ (bottom) and for two possible environments : either a uniform medium ($s=0$) with $n=A/m_\mathrm{p}=1\ \mathrm{cm^{-3}}$ (left column) or a dense stellar wind ($s=2$) with $A_{*}=A/\left(5\times 10^{11}\ \mathrm{g~cm^{-1}}\right)=0.1$ (right column).  The other parameters are fixed~:  $\tau=30\ \mathrm{s}$ which is the duration of GRB 980425 and $\kappa=4$.
The transparency constraint is dominated by equation~(\ref{eq:c2}) (optical depth for pair creation) and not by equation~\ref{eq:c1} (Thomson opacity of ambient electrons). The environment constraint (equation~(\ref{eq:c3})) is much stronger in the case of a stellar wind. For winds denser than $A_{*}=0.1$ (WR stars may have $A_{*}\ga 1$), it becomes very difficult for internal shocks to operate before deceleration starts. This problem was already mentioned in \citet{daigne:99,daigne:01}.\\

\begin{figure}
\centering
\resizebox{\hsize}{!}{\includegraphics{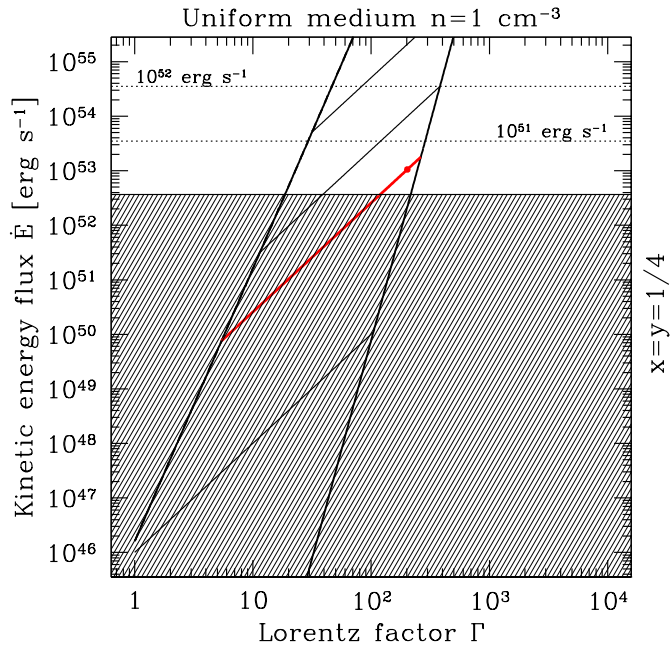}}
\caption{\textbf{Effect of the contrast $\kappa$}. The diagram at $z=1$ plotted in Figure~\ref{fig:isparameters2} has been moved to the situation where the amplitude of the initial variations of the Lorentz factor is low~: $\kappa=1.5$ instead of $\kappa=4$. All other parameters are the same as in Figure~\ref{fig:isparameters2}. The thick line corresponds to $E_\mathrm{p}=40~\mathrm{keV}$. Above this line are found $E_\mathrm{p}=70\ \mathrm{keV}$ and $150\ \mathrm{keV}$. Below is plotted the line for $E_\mathrm{p}=10\ \mathrm{keV}$.
The big dot indicates the location of a ``standard'' XRR with $E_\mathrm{p}=40~\mathrm{keV}$ (observer frame) and $L_\mathrm{rad,4\pi}=3\times 10^{50}\ \mathrm{erg~s^{-1}}$.}
\label{fig:isparameters3}
\end{figure}

\textit{Despite these constraints, the internal shock model can explain a large range of bolometric luminosities and peak energies, in agreement with the observed GRB diversity.} As can be seen in Figure~\ref{fig:isparameters1}, the internal shock model allows the existence of GRB980425-like events at $z=0.0085$, except in the case of a wind environment with $x=y=1/2$. These weak GRBs correspond to mildly relativistic ($\Gamma\simeq 10-20$) and mildly energetic ($\dot{E}\simeq 5\times 10^{47}\ \mathrm{erg~s^{-1}}$) outflows. On the other hand Figure~\ref{fig:isparameters2} confirms that at cosmological distance ($z=1$), only ``standard'' GRBs with highly relativistic ($\Gamma\ga 30$) and highly energetic ($\dot{E}\ga 10^{51}\ \mathrm{erg~s^{-1}}$ can be detected. Notice that Figure~\ref{fig:isparameters2} show no X-ray Flashes (XRFs) or X-ray rich GRBs (XRRs), i.e. no GRBs with low peak energies ($E_\mathrm{p} \la 50\ \mathrm{keV}$). As explained in \citet{barraud:05}, such soft GRBs are produced in outflows with high Lorentz factors $\Gamma$ and low contrasts $\kappa$, i.e. outflows with a very low baryonic pollution and a smoother ejection. This is illustrated in Figure~\ref{fig:isparameters3} where is shown the effect of a low constrast $\kappa$.


\subsection{The afterglow of GRB 980425-like events}


\begin{figure}
\centering
\resizebox{\hsize}{!}{\includegraphics{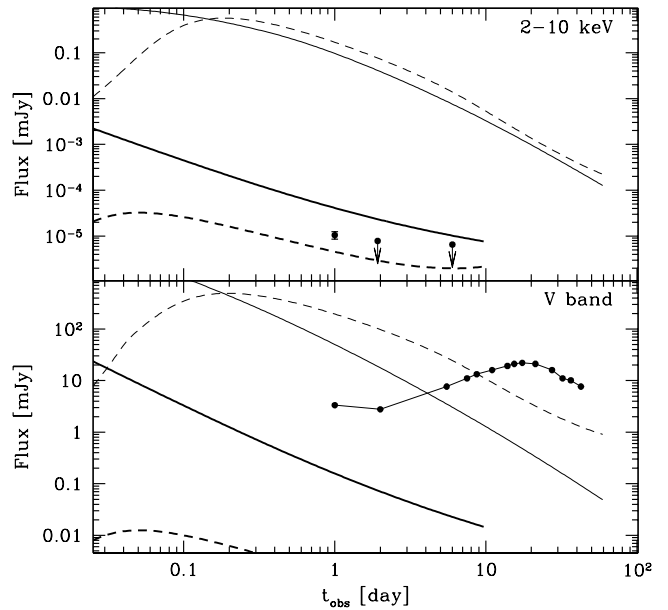}}
\caption{\textbf{Afterglow lightcurve in X-rays (top panel) and V band (bottom)}. The thick lines show the predicted afterglow lightcurve in the case of an intrinsically weak GRB seen on-axis and the thin lines in the case of a bright GRB seen off-axis. The solid lines correspond to a stellar wind environment and the dashed lines to a uniform medium (see text). Observations are shown with filled circles.}
\label{fig:afterglow}
\end{figure}

The afterglow of a ``standard'' GRB at 40 Mpc should be~$\sim 10=5\log{\left(D(z\sim 1)/40~\mathrm{Mpc}\right)}$ magnitudes brighter than for a cosmological GRB, due to the short distance. It would then easily peak at the fifth magnitude in V band. Such a bright afterglow was not detected in association with GRB 980425. The emission in V band in entirely dominated by the lightcurve of SN1998bw \citep{galama:98} (see filled circles in the bottom panel of Fig.~\ref{fig:afterglow}) and no afterglow was detected. In X-ray, the WXM of \textit{Beppo-SAX} has detected a weak decreasing source that could be the X-ray afterglow : see the filled circles in the top panel of Fig.~\ref{fig:afterglow} \citep{pian:00}. Note that only the first point is a firm detection. Any scenario for GRB 980425 should then predict a weak afterglow to be consistent with these data. We have plotted in Fig.~\ref{fig:afterglow} the predicted lightcurve in X-rays and in the V band of the afterglow for each of the two scenarios considered in this paper. Again, the external medium is either a stellar wind with $s=2$ and $A_{*}=0.1$ (solid line) or a uniform medium with $s=0$ and $n=1~\mathrm{cm^{-3}}$ (dashed line). The first case (wind environment) should be preferred due to the association of GRB 980425 with SN1998bw. The afterglow is computed assuming ``typical'' values for the parameters : $p=2.5$ (slope of the electron power law distribution), $\epsilon_\mathrm{e}=0.1$ (fraction of dissipated energy injected in relativistic electrons) and $\epsilon_\mathrm{B}=10^{-3}$ (fraction of dissipated energy injected in magnetic field). The thick lines correspond to an intrinsically weak GRB seen on-axis. According to our analysis of the prompt emission, we have adopted $\Gamma_{0}=10$ for the initial Lorentz factor and $E_{0}=6\times 10^{48}~\mathrm{erg}$ for the initial kinetic energy. The thin lines correspond to an intrinsically bright GRB seen off-axis. The initial Lorentz factor and kinetic energy have ``standard'' values $\Gamma_{0}=100$ and $E_{0}=10^{53}\ \mathrm{erg}$. The opening angle is fixed to $\Delta\theta=5^{\circ}$ and the viewing angle to $\theta_{0}=\Delta\theta+4/\Gamma_{0}$ (see section~\ref{sec:openingangle}).\\

It appears clearly that in the off-axis scenario, the predicted afterglow is much too bright. It should have been easily detected in X-rays or in the V band. This is again an evidence against the off-axis scenario \citep{waxman:04a,waxman:04b}. A discussion based on the radio afterglow leads to the same conclusion \citep{waxman:04b,soderberg:06}. To decrease the afterglow flux, one should either consider a very low density environment : $A_{*}\la 3\times 10^{-4}$ ($s=2$) or $n\la 10^{-5}\ \mathrm{cm^{-3}}$ ($s=0$), or assume very low equipartition parameters such as $\epsilon_\mathrm{e}<5\times 10^{-4}$ or $\epsilon_\mathrm{B}< 10^{-7}$. On the other hand, in the case of an intrinsically weak GRB seen on-axis, the afterglow is also very weak, due to the low value of the kinetic energy. The afterglow lightcurve in the V band is always well below the lightcurve of SN 1998bw, while in X-rays, it is very close to the \textit{Beppo-SAX} detections or upper limits.

\subsection{Rate of GRB 980425-like events}
The parameters of the internal shock scenario should in principle be given by models of the central engine. However, the production of the relativistic outflow is far from being understood. Therefore, the probability density of parameters such as the injected kinetic energy flux $\dot{E}$, the mean Lorentz $\Gamma$, the constrast $\kappa$ and the duration of the relativistic ejection $\tau$ are unknown, which prevents us from using the internal shock model to predict the rate of a specific class of GRBs, such as GRB 980425-like events. However we have shown that in the framework of the internal shock model, ``standard'' GRBs are produced by highly energetic~/~highly relativistic outflows, whereas GRB 980425-like events are produced by mildly energetic~/~mildly relativistic outflows. The fact that GRB 980425 has been observed (and maybe a somehow similar event with GRB 060218) whereas no ``standard'' GRB has been observed within a few hundreds Mpc would then indicate that less spectacular relativistic ejections are more probable than the extreme outflows that are needed to produce a ``standard'' GRB. Indeed, the fraction of stellar collapses that produce bright GRBs is estimated to be of the order of $10^{-6}$ (apparent rate) \citep{porciani:01}, i.e. $\sim 10^{-4}-10^{-3}$ after correction for the beaming angle \citep{frail:01}. Recently \citet{soderberg:06b} have estimated that the rate of ``sub-energetic GRBs'' (GRB980425-like events) has to be $\sim 10$ times larger than the rate of ``standard'' bright GRBs, which is still compatible with only a fraction of type Ic supernovae being associated with such GRBs if the beaming angle is not too small.

\subsection{Non electromagnetic emission}
As the rate of GRB 980425-like events is expected to be larger than the rate of ``standard'' GRBs, it is interesting to investigate whether such weak bursts could contribute in a significant way to the production of non electromagnetic radiation, i.e. ultra-high energy cosmic rays (UHECRs), high energy neutrinos and gravitational waves. Due to a lower Lorentz factor and a lower kinetic energy, it can be shown, using the formalism developped by \citet{waxman:01} that GRB 980425-like events are probably not able to accelerate protons at energies higher than $\sim 10^{19}~\mathrm{eV}$. In addition, the low power of these bursts is not compensated by the larger event rate so that the local energy injection rate in UHECRs due to GRB 980425-like events is probably only a few percents of the rate due to ``standard GRBs''. We conclude that GRB 980425-like events are not good candidates as sources for UHECRs, and probably do not even contribute significantly to the observed CR spectrum below $10^{19}~\mathrm{eV}$. The same kind of conclusion is reached for high-energy neutrinos. Gamma-ray photons of $\sim 100~\mathrm{keV}$ will interact with protons at the photo-meson threshold of the $\Delta$ resonance \citep[see e.g.][]{waxman:01}, i.e. of energy $\sim 2\times 10^{14}\ \mathrm{eV}$ in the observer frame for $\Gamma=10$. Due to the decay of the produced pions, neutrinos are emitted, which carry typically 5 \%
of the initial proton energy, i.e. $\sim 10^{13}~\mathrm{eV}$. However, the neutrino flux should a priori scale with the high energy proton flux, and therefore the contribution of GRB 980425-like events should be only a small fraction of the total GRB contribution.\\
The prospects are much better for gravitational waves. Long GRBs are believed to be associated with massive stars that collapse into black holes \citep{woosley:93}. As only a fraction of these collapses lead to the production of a gamma-ray burst, peculiar conditions are required. It is natural to expect that a high rotation of the progenitor favors GRBs, as the ultra-relativistic outflow is probably easier to eject along the rotation axis. Therefore, collapsars might be asymetric and then be the source of gravitational waves. This is also supported by the physical modelling of SN1998bw, that probably requires a high degree of asymmetry \citep{hoflich:99,nakamura:01}.
\citet{kobayashi:03} estimate the expected signal for an asymmetric stellar collapse to be of the order of $h\sim 10^{-22}$ in the 100-1000 Hz range at $\sim 30$ Mpc for plausible parameters. Such a signal will be detected by advanced gravitational waves detectors. It is interesting to note that in the proposed scenario the nature of the initial event responsible for GRB 980425 may not be so different than for ``standard'' GRBs. This is supported by the similarities between SN2003dh (associated with GRB 030329 at $z=0.168$) and SN 1998bw.
So the emission of gravitational waves of GRB 980425-like events may be comparable with those expected for ``standard'' long GRBs. Due to their higher rate within a few 10 Mpc, GRB 980425-like events are therefore very promising sources for advanced GW detectors \citep{norris:03}.

\section{Conclusion}
\label{sec:conclusions}
We have investigated two scenarios to explain GRB 980425-like events :\\
\noindent -- in the first scenario, GRB 980425 is a normal (intrinsically bright) GRB seen off-axis. A viewing angle that is greater than the opening angle of the jet by a few $1/\Gamma$ is enough to produce an apparently weak GRB like 980425. However this scenario suffers several problems : (1) the afterglow of GRB 980425 should have been very bright and therefore detected; (2) the rate of GRB 980425-like events should be very low. 
If we exclude the possibility that we observed by chance an extremely rare event and if we take into account the fact that a local on-axis intrinsically bright GRB has never been observed, we get two strong constraints : the local rate of ``standard'' bright GRBs has to be much higher than what can be simply extrapolated from the cosmic event rate and at the same time the typical jet opening angle in GRBs must be much narrower than what is usually inferred from observations of breaks in afterglow lightcurves.\\
\noindent -- \textit{Since this appears rather unrealistic, we prefer the second scenario, where GRB 980425 is an intrinsically low-luminosity GRB seen on-axis.} We have shown that the parameter space of the internal shock model allows such events, that would be produced by mildly relativistic ($\Gamma\sim 10-20$), mildly energetic ($\dot{E}\simeq 5\times 10^{47}\ \mathrm{erg.s^{-s}}$) outflows. The consequences are that (1) in collapsars, the central engine in most cases fails to power a highly-relativistic/highly-energetic ejection (necessary for bright ``standard'' GRBs) but can more easily power a mildly relativistic/mildly-energetic outflow; (2) the rate of GRB 980425-like intrinsically weak events in the Universe is then much higher than the rate of ``standard'' intrinsically bright GRBs but the apparent rate is lower because these events can be detected only at low redshift. Finally, as this scenario infers a higher local rate for GRB 980425-like events compared to ``standard'' bright GRBs, we have discussed the possible non-electromagnetic emission of these bursts. We find that they are not promi\-sing sources of UHECRs and high-energy neutrinos, as the corresponding flux probably scales with the gamma-ray flux. On the other hand, 
since GRB 980425 was associated with an extreme type Ic supernova as those associated with ``standard'' long GRBs, 
one can expect the corresponding stellar collapse to be asymmetric and GRB 980425-like events could therefore be promising sources of gravitational waves within a few 10 Mpc.

\bibliographystyle{aa}
\bibliography{weakgrb}

\end{document}